\documentclass[12pt]{article}

\usepackage{cite}
\usepackage{graphicx,color,epsfig,rotating}
\usepackage{amsfonts,amsmath,amssymb,bbm}
\usepackage{algorithm}
\usepackage{algpseudocode}
\usepackage{subfigure}
\usepackage{amsmath}
\usepackage{cite}
\usepackage{placeins}
\usepackage{graphicx}
\usepackage[latin1]{inputenc}
\usepackage{amssymb}
\usepackage{multirow}
\usepackage{stfloats}
\usepackage{tabularx} 
\usepackage{booktabs} 
\usepackage{url}
\usepackage{bm}
\usepackage{soul}
\usepackage{float}
\usepackage{lipsum}     
\usepackage{cuted} 
\usepackage{pstricks}
\usepackage{arydshln}
\usepackage{xspace} 
\usepackage{array}
\usepackage{booktabs}
\usepackage{mathtools}
\usepackage{authblk}
\allowdisplaybreaks  
\long\def\comment#1{}

\makeatletter
\newcommand{\subalign}[1]{
  \vcenter{%
    \Let@ \restore@math@cr \default@tag
    \baselineskip\fontdimen10 \scriptfont\tw@
    \advance\baselineskip\fontdimen12 \scriptfont\tw@
    \lineskip\thr@@\fontdimen8 \scriptfont\thr@@
    \lineskiplimit\lineskip
    \ialign{\hfil$\m@th\scriptstyle##$&$\m@th\scriptstyle{}##$\crcr
      #1\crcr
    }%
  }
}

\newcommand{\thickhline}{%
    \noalign {\ifnum 0=`}\fi \hrule height 1pt
    \futurelet \reserved@a \@xhline
}
\newcolumntype{"}{@{\hskip\tabcolsep\vrule width 1pt\hskip\tabcolsep}}

\makeatother

\newtheorem{example}{Example}
\newtheorem{theorem}{Theorem}
\newtheorem{remark}{Remark}

\newcommand{\xz}[1]{{\color{red} [XZ's comment: #1]}}

\def \Utwoone{{U_2^{(1)}}}

\def \rx{{R_X}}

\def \rxstar{{R_X^*}}

\let \mrm  \mathrm
\let\trm\textrm
\let\tbf\textbf
\let\tit\textit

\let\mbb\mathbb

\let \bksl\backslash

\let \udb \underbrace

\newcommand{\user}[1]{user~$#1$}

\newcommand{\User}[1]{User~$#1$}

\newcommand{\nece}{necessary\xspace}

\newcommand{\corrd}{correlated\xspace}

\newcommand{\recov}{recover\xspace}

\newcommand{\recovy}{recovery\xspace}

\newcommand{\Recovy}{Recovery\xspace}

\newcommand{\compo}{component\xspace}
\newcommand{\compos}{components\xspace}

\newcommand{\pk}{pairwise key\xspace}
\newcommand{\pks}{pairwise keys\xspace}

\newcommand{\PKs}{Pairwise Keys\xspace}

\newcommand{\oppo}{opposite\xspace}
\newcommand{\rcvd}{received\xspace}

\newcommand{\eqvlt}{equivalent\xspace}
\newcommand{\eqvlce}{equivalence\xspace}

\newcommand{\arbicorr}{arbitrarily correlated\xspace}

\newcommand{\Secty}{Security\xspace}
\newcommand{\secty}{security\xspace}

\newcommand{\ntwk}{network\xspace}

\newcommand{\exendmark}{\hfill$\lozenge$}

\newcommand{\nbrs}{neighbors\xspace}
\newcommand{\nbr}{neighbor\xspace}

\newcommand{\nbhd}{neighborhood\xspace}
\newcommand{\Nbhd}{Neighborhood\xspace}

\newcommand{\arbily}{arbitrarily\xspace}
\newcommand{\arbi}{arbitrary\xspace}

\newcommand{\Thf}{Therefore\xspace}

\newcommand{\Aar}{As a result\xspace}

\newcommand{\brdcst}{broadcast\xspace}

\newcommand{\decen}{decentralized\xspace}

\newcommand{\Inadd}{In addition\xspace}

\newcommand{\genn}{generation\xspace}
\newcommand{\Genn}{Generation\xspace}

\newcommand{\topo}{topology\xspace}

\newcommand{\topos}{topologies\xspace}
\newcommand{\topocl}{topological\xspace}

\newcommand{\charz}{characterize\xspace}

\newcommand{\eg}{e.g.\xspace}

\newcommand{\ie}{i.e.\xspace}
\newcommand{\msg}{message\xspace}

\newcommand{\msgs}{messages\xspace}

\newcommand{\hie}{hierarchical\xspace}

\newcommand{\specly}{specifically\xspace}

\newcommand{\Ip}{In particular\xspace}

\newcommand{\af}{as follows\xspace}
\newcommand{\resp}{respectively\xspace}
\newcommand{\iid}{i.i.d.\xspace}
\newcommand{\iidu}{i.i.d. uniform\xspace}

\newcommand{\Thm}{Theorem\xspace}
\newcommand{\schm}{scheme\xspace}
\newcommand{\schms}{schemes\xspace}

\newcommand{\info}{information\xspace}

\newcommand{\etal}{\textit{et al.}\xspace}

\newcommand{\bcuz}{because\xspace}
\newcommand{\regu}{regular\xspace}

\newcommand{\agg}{aggregation\xspace}

\newcommand{\secagg}{secure aggregation\xspace}
\newcommand{\Secagg}{Secure aggregation\xspace}

\newcommand{\diff}{different\xspace}

\newcommand{\Diff}{Different\xspace}

\newcommand{\Fex}{For example\xspace}

\newcommand{\indep}{independent\xspace}

\newcommand{\indepce}{independence\xspace}

\newcommand{\indiv}{individual\xspace}

\newcommand{\comm}{communication\xspace}

\newcommand{\Comm}{Communication\xspace}

\newcommand{\achved}{achieved\xspace}
\newcommand{\achvb}{achievable\xspace}

\newcommand{\distn}{distribution\xspace}

\newcommand{\dist}{distance\xspace}
\newcommand{\muinfo}{mutual information\xspace}

\newfont{\bbb}{msbm10 scaled 700}

\newfont{\bb}{msbm10 scaled 1100}

\newcommand{\Ac}{{\cal A}}
\newcommand{\Bc}{{\cal B}}

\newcommand{\Ec}{{\cal E}}

\newcommand{\Gc}{{\cal G}}

\newcommand{\Nc}{{\cal N}}

\newcommand{\Sc}{{\cal S}}

\newcommand{\Vc}{{\cal V}}
\newcommand{\Xc}{{\cal X}}

\newcommand{\Zc}{{\cal Z}}

\newcommand{\inputsumnbrk}{\hbox{$\sum_{i\in \Nc_k}W_i$}}

\newcommand{\eqdef}{\stackrel{\Delta}{=}}

\newcommand{\be}{\begin{equation}}
\newcommand{\ee}{\end{equation}}
\newcommand{\bea}{\begin{eqnarray}}
\newcommand{\eea}{\end{eqnarray}}

\newcommand{\besub}{\begin{subequations}}
\newcommand{\eesub}{\end{subequations}}

\newcommand{\BLUE}{\color[rgb]{0,0,0.90}}

\usepackage[left=2cm,right=2cm, top=.77in, bottom=1.12in]{geometry} 

\begin{document}

\title{Optimal Communication Rate of Secure Aggregation over Ring Networks with Pairwise Keys}

\author[1]{Xiang Zhang}
\author[1]{Han Yu}
\author[2]{Zhou Li}
\author[3]{Yizhou Zhao}
\author[1]{Giuseppe Caire}

\affil[1]{Department of Electrical Engineering and Computer Science, Technical University of Berlin

\texttt{Email:\{xiang.zhang,caire\}@tu-berlin.de,~han.yu.1@campus.tu-berlin.de}
}

\affil[2]{Guangxi Key Laboratory of Multimedia Communications and Network Technology, Guangxi University

\texttt{Email:lizhou@gxu.edu.cn}
}

\affil[3]{College of Electronic and Information Engineering, Southwest University

\texttt{Email:onezhou@swu.edu.cn}}

\renewcommand\Authfont{\normalsize}
\renewcommand\Affilfont{\small}


\maketitle

\begin{abstract}
Information-theoretic topological secure aggregation (TSA)\cite{zhang2026information_regular} enables distributed users to compute neighborhood sums over arbitrary networks without revealing individual inputs, while remaining communication-efficient. It has broad applications, including secure model aggregation in decentralized federated learning (FL). 
Existing TSA formulations rely on arbitrarily correlated keys generated by a trusted key server, which introduces a single point of failure.
In this paper, we instead study TSA with \tit{pairwise} secret keys, where each user pair $(i,j)$ shares an independent key $S_{i,j}$.
Such keys can be established through inter-user communication, eliminating the need for a key server and improving robustness. 
Focusing on a ring topology with $K$ users, we characterize the minimum per-user communication rate: \tit{to securely  compute  one bit of the  desired input sum, each user must send at least $1$ bit to its neighbors when $K=3,4$, and at least $2$ bits for all $K\ge 5$}. 
The higher rate in larger networks arises because each user must simultaneously satisfy two independent key-alignment constraints from its two neighborhoods, which cannot be resolved within a single broadcast symbol under pairwise key independence.
We propose a linear pairwise-masking scheme that achieves these rates and prove its optimality via tight entropic converse bounds that exploit the dependency structure of the keys.
Notably, for all $K\ge 4$, only a subset of the $\binom{K}{2}$ pairwise keys---specifically, those
between users at ring distance $2$---is sufficient to achieve optimality, revealing a nontrivial role of topological sparsity in secure aggregation.
\end{abstract}

\section{Introduction}
\label{sec:intro} 

Federated learning (FL)~\cite{mcmahan2017communication} enables collaborative training of machine learning models  on distributed datasets without sharing clients' raw datasets. In the centralized setting,  an  \agg server coordinates the  aggregation process in a star \ntwk; in decentralized FL (DFL)~\cite{kairouz2021advances,beltran2023decentralized}, aggregation is performed across neighboring nodes over a communication graph.
Despite keeping datasets local, FL remains vulnerable when the server or peers are untrusted~\cite{bouacida2021vulnerabilities}. This has driven extensive work on secure aggregation (SA) in FL~\cite{bonawitz2017practical,wei2020federated,hu2020personalized}, where cryptographic tools are used to ensure  the \secty of the local datasets.
For instance, Bonawitz \etal~\cite{bonawitz2017practical} introduced a protocol based on pairwise secret-key masking to protect local models from an untrusted server. Since then, numerous SA schemes built on cryptographic primitives have been proposed, largely emphasizing \tit{computational} security.

\Diff from cryptography-based SA,
information-theoretic secure aggregation (IT-SA)\cite{9834981, 10359136,aggregation_light,zhao2023secure,so2022lightsecagg,li2023weakly,zhang2025optimal,zhang2025fundamental,chou2026private,zhang2026information} 
ensures \tit{perfect} security by enforcing strict statistical independence (via zero mutual information) between the masked and true model updates.  It offers principled guidance for designing provably secure and communication-efficient FL protocols.  
IT-SA has been studied under various constraints, including user dropout resilience~\cite{9834981, zhao2023secure,so2022lightsecagg,jahani2023swiftagg+}, groupwise keys~\cite{zhao2023secure,wan2024information,li2025capacity}, heterogeneous security constraints~\cite{li2023weakly,li2025weakly}, oblivious server~\cite{sun2023secure},  multiple recovery objectives~\cite{yuan2025vectorlinearsecureaggregation,hu2026capacity}, leakage SA\cite{chou2026private}, \hie SA (HSA)~\cite{zhang2024optimal, lu2024capacity,zhang2025fundamental,egger2025private,xu2025hierarchical}, and \decen SA (DSA)\cite{zhang2026information,li2025capacity}.

One important branch of IT-SA is \tit{\topocl SA (TSA)}\cite{zhang2026information_regular,zhang2026information}, which studies \decen \secagg in \arbi \comm \topos.  \Ip, \cite{zhang2026information} investigated SA in a fully connected \topo where each user wants to recover the sum of all users' inputs (abstraction of local updates in FL). The optimal \comm and secret key rates were  shown equal to one, which is universally minimal, due to the dense connectivity of the \ntwk. \cite{zhang2026information_regular} considered several types of $d$-\regu sparse networks and showed that the optimal \comm and key rates depend on the kernel size of a diagonally modulated adjacency matrix of the  \comm graph, revealing a nontrivial impact of \topo on SA.
However, both works\cite{zhang2026information_regular,zhang2026information} assume the users' keys can be \arbicorr  and have to rely on a trusted key server for key \genn and \distn. The dedicated key server not only adds a new attack  front, but also incurs global \comm overhead due to key \distn.

\tbf{Contribution.}
In this work, we study TSA with \tit{pairwise} keys, where each pair of users $(i,j)$ shares an \indep key $S_{i,j}$. 
Such keys can be established via inter-user communication, eliminating the need for a dedicated key server and improving robustness.
We first formulate TSA with \pks  over a general \comm graph $\Gc=(\Vc, \Ec)$ with vertex  and edge sets $\Vc(|\Vc|=K)$ and $\Ec$ representing  the users and inter-user connections, \resp. 
Through \nbhd broadcast transmissions,
each  user aims to securely compute the sum of  its \nbrs' private inputs, while being  prevented from  learning the \indiv inputs. 
Then, for the ring \topo commonly used in DFL\cite{dai2022dispfl}, we \charz the optimal  \comm rate: \tit{to securely compute one bit of the desired sum at each user, each user needs to broadcast at least $1$ bit when $K =3,4$, and at least $2$ bits when $K\ge 5$.} 
This result is achieved via a novel input masking scheme that selectively uses a strict subset of the  $\binom{K}{2}$ \pks, 
together with an entropic converse proof establishing tight lower bounds on the \achvb \comm rates. 
Overall, our result reveals a fundamental interplay between network topology and key structure in determining communication efficiency.

\emph{Notation.} 
$[n] \eqdef \{1,  \cdots, n\} $,
$\Ac\bksl \Bc  \eqdef \{x\in  \Ac:  x \notin  \Bc \}$. 
$\binom{\Ac}{n} \eqdef \{\Sc \subseteq \Ac: |\Sc|=n\}$, and $X_\Ac \eqdef \{X_i\}_{i\in \Ac} $. 

\section{Problem Statement}
\label{sec:problem statement}
Consider  a  \comm  network represented  by  connected and undirected graph $\Gc=(\Vc, \Ec)$, where the vertex set $\Vc $ represents the $K \ge  3$ users and the edge set $\Ec$ represents the undirected \comm links  among them.
Let $\Nc_k \subseteq [K]\bksl \{k\} $  denote the neighbor set of user $k$. 
User $k$ holds an \emph{input} variable $W_k \in \mbb{F}_q^{1\times L}$ which
consists  of $L$ \iid uniform bits, representing locally trained model/gradient updates in federated learning. 
The inputs of \diff users are assumed to be \indep\footnote{The proposed \secagg scheme works for \arbily distributed and correlated inputs. The
\iidu assumption is only necessary to establish the optimality of the scheme.}, \ie,
\be
\label{eq:input indep}
H(W_k)  = L, \; k \in [K], \quad
H(W_{1:K})  = \sum_{k=1}^K H(W_k).
\ee

\tbf{Pairwise key model.}
Let $S_{i,j}$ be a secret key shared exclusively by users $i$ and $j$ with  $j>i$. For ease of presentation, let $S_{i,j}=-S_{j,i}$ if $ i >j$, \ie,
$S_{i,j}$ and $S_{j,i}$ are essentially the same random variable up to a sign change.
The keys are mutually \indep and also \indep of the inputs, 
\be
\label{eq:key-key,key-input indepce}
H\left(\{S_{i,j}\}_{i<j},W_{1:K}\right)= \sum_{i<j}H(S_{i,j}) + \sum_{k=1}^KH(W_k).
\ee 
\User{k} then stores 
\be 
\label{eq:Zk}
\Zc_k = \{S_{i,k}\}_{i<k} \cup \{S_{k,j}\}_{j>k},  \; k \in[K]
\ee
Note that although $\Zc_k$ contains all $\binom{K}{2}$ possible \pks, a specific  \secagg \schm may only use a subset of them as we will see later. 

\begin{remark}[Key \Genn]
\label{rmk:key gen}
\tit{Pairwise keys can be efficiently established via standard cryptographic primitives (\eg, Diffie-Hellman\cite{maurer2000diffie}). The key-establishment topology need not match the aggregation graph $\Gc$.
Here we assume a fully connected key-establishment model where $S_{i,j}$ exists for any user pair $(i,j)$. 
Since keys are generated offline and reused across rounds, the one-time setup cost is thus amortized and negligible.}
\end{remark}

\tbf{\Comm protocol.}
Each user is connected to its neighbors through an error-free broadcast channel, and the channels originating  from  \diff users are assumed to be orthogonal. 
Each user aims to recover the sum of the inputs of its \nbrs. To this end, each \user{k} generates a \msg $X_k$, which is broadcast to all users in $\Nc_k$. $X_k$ is
a deterministic function of the input $W_k$ and key $Z_k$, 
\be
\label{eq:msg gen}
H(X_k|W_k, \Zc_k)=0.
\ee 
After receiving the \msgs from all its \nbrs, each user should correctly \recov  the  sum  of their inputs (including itself), \ie,  
\begin{align}
\label{eq:recovery} 
H\left(\sum\nolimits_{i\in \Nc_k \cup \{k\}  }W_i \Big|(X_i)_{i \in \Nc_k}, W_k, \Zc_k  \right)=0, \;\forall k\in[K]
\end{align}
Note that recovering $\inputsumnbrk$  suffices to ensure the above recovery constraint since $W_k$ belongs to  user $k$.

\tbf{Security model.}
We consider the \emph{honest-but-curious} model commonly used in secure FL and  multi-party computation (MPC)~\cite{bonawitz2017practical, goldreich1998secure,jahani2023swiftagg+}, where each user executes the aggregation protocol faithfully but attempts to infer its neighbors'  private inputs from the received \msgs.
\Thf,  \secty  is imposed \tit{against} each user:  each \user{k} must not infer any information about the \nbhd inputs $(W_i)_{i\in \Nc_k}$ beyond their  sum $\sum_{i \in \Nc_k}W_i$, which can be expressed as
\begin{align}
\label{eq:security}
& I\left((X_i)_{i\in \Nc_k  }   ;(W_i)_{i\in \Nc_k} \Big| \sum\nolimits_{i\in \Nc_k}W_i , W_k, \Zc_k \right)=0. 
\end{align}
The conditioning terms arise from the fact that user $k$ must eventually recover $\sum_{i\in \Nc_k}W_i$.

The \comm rate $\rx$ is defined as the normalized (by input size) entropy of each \brdcst \msg, 
\be
\label{eq:def Rx}
\rx \eqdef H(X_k)  /L.
\ee
We aim to find the minimum \achvb \comm rate $\rxstar$ over all \secagg \schms.

\section{Main Result}
\label{sec:main result}

\begin{theorem}
\label{thm:main thm}
For  \secagg  over the $K$-user ring \ntwk with \pks, 
 the optimal \comm rate is given by
\be 
\label{eq:R*,thm}
\rxstar=
\begin{cases}
1, & \mathrm{if}\; K =3,4\\
2, & \mathrm{if}\; K\ge 5
\end{cases}
\ee
\end{theorem}

\tit{Proof:}
The proof of \Thm \ref{thm:main thm} consists of an \achvb  \schm and a matching converse proof presented  in Sections \ref{sec:proposed scheme} and \ref{sec:converse}, \resp. 
The  results can be interpreted \af:

\tit{1) Optimal Rate:}
The key intuition behind the one-to-two transition of $\rxstar$ at $K=4,5$ lies in  \pk alignment.
To let user \(k\) securely recover
\(W_{k-1}+W_{k+1}\), the two neighboring messages $X_{k-1},X_{k+1}$ must contain  $1$ 
bit (suppose $L=1$) of correlated randomness that cancel at user \(k\) but remain
unknown to it. 
Under pairwise keys, this role is naturally played by
\(S_{k-1,k+1}\), so \(X_{k-1}\) and \(X_{k+1}\) must carry this key with
opposite signs.
Therefore, $X_k$ must support two distinct
recoveries: one aligned with \(S_{k-2,k}\) for user \(k-1\) to \recov $W_{k-2} +W_k$,
and the other
with \(S_{k,k+2}\) for user \(k+1\) to \recov 
 $W_{k} +W_{k+2}$. 
For $K=3$, 
the ring degenerates into a fully connected network where  $\rxstar=1$ as shown by \cite{li2025capacity}. 
For $K=4$, both the recoveries  at users $2$ and $3$ require the use of $S_{1,4}$, so that a single masked  input by $S_{1,4}$ suffices, resulting in $\rxstar=1$. 
However,  when $K\ge  5$,  the two keys  
 \(S_{k-2,k}\) and \(S_{k,k+2}\) are \indep, and
neither can be canceled by the unintended receiver since  \(S_{k-2,k} \notin \Zc_{k+1} \),
\(S_{k,k+2} \notin \Zc_{k-1} \). 
Thus, they cannot be merged into a single masked input. Two independent input dimensions are therefore necessary to support the two recovery constraints at users $k-1$ and $k+1$, yielding $\rxstar=2$.

\tit{2) Impact of \PKs:} \Secagg over ring \topos  has also been studied under a
\diff secret key model \cite{zhang2026information_regular}, where user keys can be \tit{\arbi} \corrd and the optimal rate is $\rx=1$ for all $K$. 
In contrast, the structural constraint imposed by pairwise keys limits such correlation, leading to an increased communication rate when $K\ge 5$.


\section{Proposed Scheme}
\label{sec:proposed scheme}

\subsection{Motivating Examples}
\label{subsec:examples}
\begin{example}[$K=3$]
\label{exam:K=3}
The $3$-user ring is a fully-connected \topo, where the optimal rate was given by Li \etal\cite{li2025capacity} as $\rxstar=1$. For completeness, we  include this example here. Suppose each input has $L=1$ bit. Let $S_{1,2}, S_{1,3}$ and $S_{2,3}$ be three \iidu bits. The \msgs are 
$
X_1= W_1+S_{1,2} + S_{1,3}, X_2= W_2+S_{2,1} + S_{2,3} 
$, and $X_3= W_3+S_{3,1} + S_{3,2}$, where recall that $S_{i,j}=-S_{j,i}$. Each user sums up all three  \msgs to  \recov $W_1+W_2+W_3$ as the keys cancel out pairwisely:  $X_1+X_2+X_3=W_1+W_2+W_3 +\sum_{i\ne j}(S_{i,j} + S_{j,i})=W_1+W_2+W_3$. \Secty is also ensured \bcuz from each  \user{k}'s view, the two \rcvd \msgs are masked by an unknown key $S_{\{1,2,3\}\bksl k}$. \Fex, after removing $S_{2,1}$ and $S_{3,1}$, \user{1} sees $W_2 + S_{2,3} $ and  $W_3+ S_{3,2}$. Besides  $W_2+W_3$, \user{1} cannot infer any \info about $(W_2,W_3)$.
\exendmark
\end{example}

\begin{example}[$K=4$]
\label{exam:K=4}
Consider the $4$-user ring 
with  \nbr sets $\Nc_1=\Nc_4=\{2,3\}$, $\Nc_2=\Nc_3=\{1,4\}$ as shown  in Fig. \ref{fig:ring 4,exam}.
\begin{figure}[ht]
    \centering
    \includegraphics[width=0.3\linewidth]{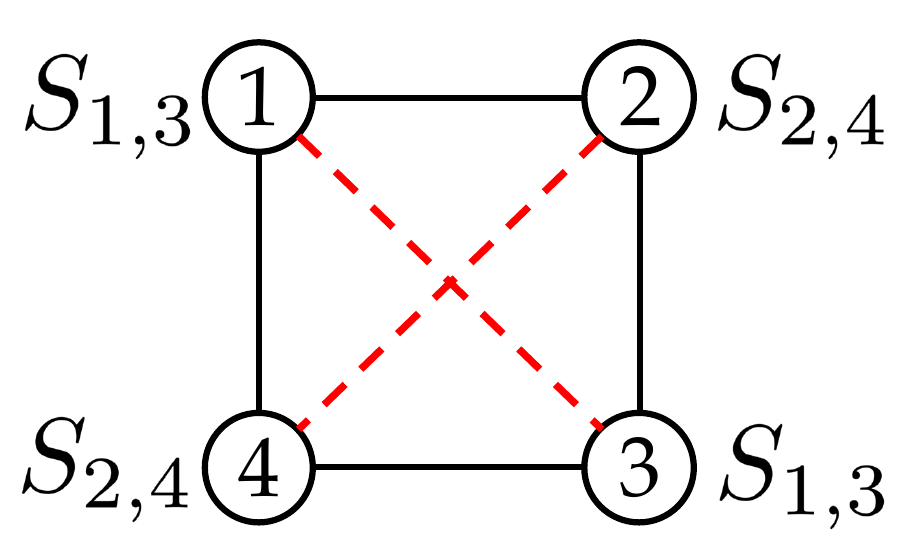}
    \caption{\small Key assignment  for $4$-user ring. Only two keys $S_{1,3}$ and $S_{2,4}$ (indicated by red dashed lines) out of the six \pks are used. }
    \label{fig:ring 4,exam}
\end{figure}
Suppose each input has one bit. Let $S_{1,3}$ and $S_{2,4}$ be two \iidu bits, and set 
$S_{i,j}=0$  for all  other $i<j$.
Hence, the keys of the users are
$
\Zc_1 =\Zc_3 = \{ S_{1,3}\}, 
\Zc_2 =\Zc_4 = \{ S_{2,4}\}
$.
The \msgs are
\begin{align}
X_1 &= W_1 +S_{1,3}, \; X_2 = W_2 +S_{2,4}, \notag\\
X_3 &= W_3 +S_{3,1}, \; X_4 = W_4 +S_{4,2}.
\end{align}

\tbf{Recovery.}
From each user's view, the inputs contained in the two \rcvd \msgs are masked by the same \pk (with \oppo signs) that is not stored by the user.
A direct sum of the two \msgs yields the desired input sum.
\Fex, \user{1}  sees  $X_2=W_2 + S_{2,4}$ and $X_4=W_4+ S_{4,2}$, from which 
$W_2+ W_4$  can be  recovered from $X_2+  X_4=  (W_2 + S_{2,4}) + (W_4+ S_{4,2})=W_2+ W_4$. Since the $4$-user ring is highly symmetric, \user{3}'s \recovy is the same as \user{1}'s,  while users $2$ and $4$ are also the same.  
Since each user  broadcasts one bit, $\rx=1$.
\exendmark
\end{example}

\begin{example}[$K=5$]
\label{exam:K=5}
Consider the $5$-user ring as shown in Fig. \ref{fig:ring 5,exam}.
\begin{figure}[ht]
\centering
\includegraphics[width=0.3\linewidth]{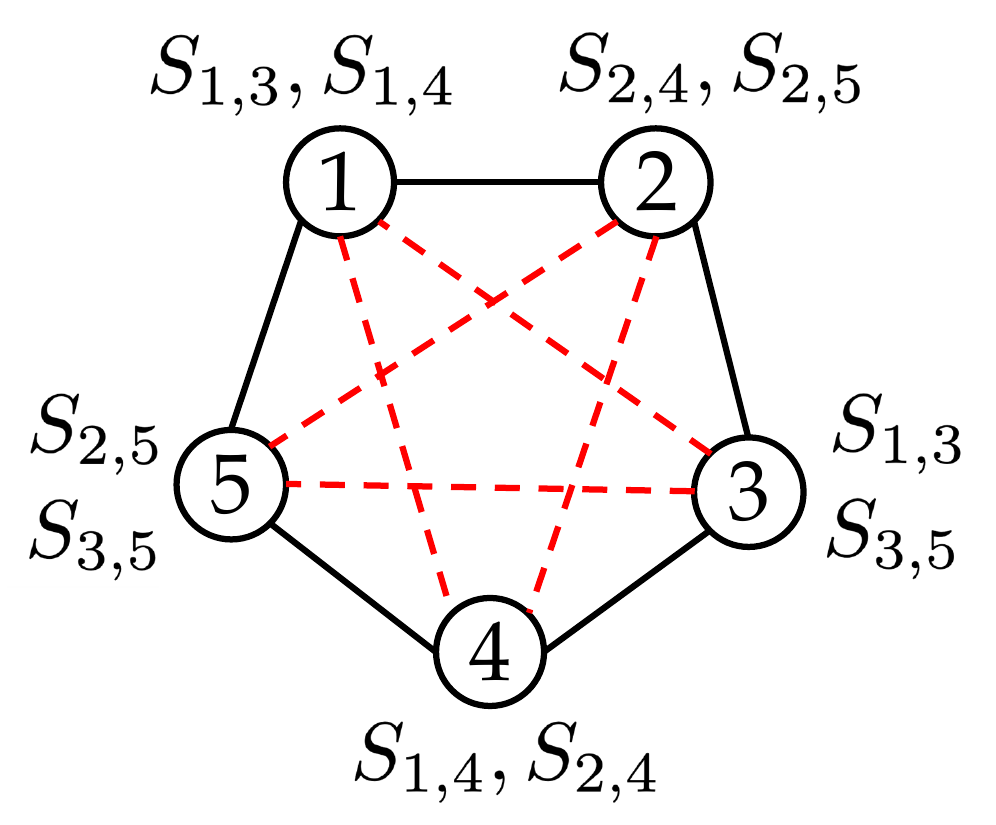}
\caption{\small Key assignment for $5$-user ring. Five \pks 
$S_{1,3}, S_{2,4}, S_{3,5}, S_{1,4} $ and $S_{2,5}$ are used. }
\label{fig:ring 5,exam}
\end{figure} 
Let $L=1$ bit.
Only keys shared by users at ring distance $2$ are  used: \specly, let $S_{1,3}, S_{2,4}, S_{3,5}, S_{1,4} $ and $S_{2,5}$ be five \iidu bits.
The \msgs, each containing two bits, are
\begin{align}
\label{eq:msgs,K5 ring,exam}
X_1 &=(X_1^{(2)}, X_1^{(  5)})
=(W_1+S_{1,3},W_1+S_{1,4}), \notag\\
X_2 &=(X_2^{( 3)}, X_2^{(  1)} )
=(W_2+S_{2,4},W_2+S_{2,5}), \notag\\
X_3 &=(X_3^{( 4)}, X_3^{( 2)} )
=(W_3+S_{3,5},W_3+S_{3,1}), \notag\\
X_4 &=(X_4^{( 5)}, X_4^{( 3)} )
=(W_4+S_{4,1},W_4+S_{4,2}), \notag\\
X_5 &=(X_5^{( 1)}, X_5^{( 4)} )
=(W_5+S_{5,2},W_5+S_{5,3}), 
\end{align}
where  $X_i^{(j)}$ denotes the \msg \compo of $X_i$ intended  for \agg  at \user{j\in \Nc_i}.
Since each  \msg has two bits, the \achved rate  is $\rx =2$.

\tbf{Recovery.}
\User{k} sums up the two  \msg  \compos intended at itself from the two \rcvd \msgs to \recov the desired input sum. \Ip, 
\begin{align}
\label{eq:recovery,K5 example}
& \trm{\User{1}:}\;  
X_5^{(1)} + X_2^{(1) } =  W_5 + S_{5,2}+W_2+ S_{2,5} =W_2 +W_5,\notag\\
& \trm{\User{2}:} \;
X_1^{(  2)} + X_3^{(  2)} = W_1 + S_{1,3} +W_3+ S_{3,1} =W_1 +W_3,  \notag\\
& \trm{\User{3}:} \; 
X_2^{( 3)} + X_4^{( 3)} = W_2 + S_{2, 4} +W_4 + S_{4,2} =W_2 +W_4,  \notag\\
& \trm{\User{4}:}\;  
X_3^{( 4)} + X_5^{( 4)} = W_3 + S_{3, 5} +W_5+S_{5,3} =W_3 +W_5,  \notag\\
& \trm{\User{5}:}\;  
X_4^{( 5)} + X_1^{( 5)} = W_4+S_{4,1} +W_1+ S_{1,4} =W_1 +W_4. \notag
\end{align}

\tbf{\Secty.}
Besides the two \msg \compos  desired for \agg, \user{k} also observes two additional \compos. The \msg design in \eqref{eq:msgs,K5 ring,exam} ensures  that i) the two desired \compos are masked by the same  \pk with opposite signs, and ii) the two additional \compos are each protected by an \indep key, which together guarantees  \secty.
\Fex, consider \user{1}, which observes
$X_2^{( 1)}=W_2+ S_{2,5},X_5^{( 1)}=W_5+ S_{5,2}, X_2^{( 3)}=W_2+ S_{2,4}$ and $X_5^{( 4)}=W_5+ S_{5,3}$. Besides the cancellation of $S_{2,5}$ and $S_{5,2}$ which yields $W_2+W_5$, $X_2^{( 3)}$ and $X_5^{( 4)}$ are  \resp masked by $S_{2,4}$ and $S_{5,3}$,  both of which are not contained in $\Zc_1$. \Thf, \secty at \user{1} is guaranteed. 
More formally, 
\begin{subequations}
\begin{align}
&I (X_2,X_5;W_2, W_5|W_2+W_5,W_1, \Zc_1)\\ &\overset{\eqref{eq:key-key,key-input indepce}, \eqref{eq:msg gen}}{=}
I (X_2,X_5;W_2, W_5|W_2+W_5, \Zc_1)\\
& = \udb{H(X_{\{2,5\}}|W_2+W_5, \Zc_1) }_{\rm Term\; 1} 
- \udb{H(X_{\{2,5\}}|W_2,W_5, \Zc_1)}_{\rm Term \;2}. 
\label{eq01}
\end{align}
\end{subequations}
The two terms in \eqref{eq01} are calculated \af:
\begin{subequations}
\begin{align}
\mrm{Term\;2} & = H(X_2, X_5|W_2,W_5, \Zc_1)\\
&  \overset{\eqref{eq:msgs,K5 ring,exam}}{=}
H(S_{2,4}, S_{2,5}, S_{5,2}, S_{5,3}| W_2,W_5, \Zc_1      )\\
& \overset{\eqref{eq:key-key,key-input indepce}}{=}
H(S_{2,4}, S_{2,5}, S_{5,3} )\label{eq002}\\
&= H(S_{2,4})+ H(S_{2,5})  + H(S_{5,3})=3, 
\end{align}
\end{subequations}
where \eqref{eq002} is due  to the  \indepce among the inputs and keys.
The first term is upper bounded by
\begin{subequations}
\begin{align}
\mrm{Term\;1} &= H(X_2, X_5|W_2+W_5, \Zc_1)\\
 & \overset{\eqref{eq:msgs,K5 ring,exam},\eqref{eq:key-key,key-input indepce}}{=}
 H(X_2, X_5|W_2+W_5)\label{eq0015}\\
 &=  H(X_2, X_5, W_2+W_5) -  H(W_2+W_5)\\
 &= H(X_2^{ (1)},X_5^{ (1)}, X_2^{ (3)}, X_5^{ (4)}, W_2+W_5   )-1\\
 & \overset{\eqref{eq:msgs,K5 ring,exam}}{=} H (X_2^{ (1)},X_5^{(1)}, X_2^{(3)}, X_5^{ (4)})-1\label{eq0002}\\
& \le 3, \label{eq0003}
\end{align} 
\end{subequations}
where  \eqref{eq0015}  is  due to the \indepce between $\Zc_1$ and $(X_2, X_5,W_2+W_5)$;
\eqref{eq0002} is \bcuz $ X_2^{(1)} + X_5^{(1)}=W_2+W_5$; \eqref{eq0003} is \bcuz $H(X_2^{ (1)},X_5^{(1)}, X_2^{ (3)}, X_5^{(4)} )\le 4$ since each \msg \compo has one bit and  uniform \distn maximizes entropy. 
\Aar,  $\eqref{eq01}=\mrm{Term\;1} -\mrm{Term\;2}\le 0$. Since \muinfo is nonnegative, we have $I (X_2,X_5;W_2, W_5|W_2+W_5,W_1, \Zc_1)=0$, proving \secty at \user{1}.
By symmetry, the proof is similar for other users.
\exendmark
\end{example}

\subsection{General Scheme}
\label{subsec:gen scheme} 
Consider $K \ge 5$.
For notational convenience, let us temporarily relabel  the users as $0, \cdots, K-1$, where all index operations are modulo $K$. 
Hence, \user{k}'s \nbr set is $\Nc_k=\{k-1, k+1\}$.  
Let  $S_{0,2}, S_{1,3},\cdots, S_{K-2,0}, S_{K-1,1}$ be $K$ \iidu bits,  and set $S_{i,j}=0$ for all other $(i,j)$. 
Thus,  only the \pks  with  ring  \dist $2$ are used. 
Suppose each input  has $L=1$ bit.
Each user sends a $2$-bit \msg
\be
\label{eq:msg Xk,gen schm}
X_k = \big(X_k^{(k-1)}, X_k^{(k+1)} \big),\; k \in [0:K-1]
\ee 
where  $X_k^{(j)}$ denotes the \msg \compo intended for \agg at \user{j \in \Nc_k}, and 
\be
\label{eq:msg Xk compos,gen schm}
X_k^{(k-1)} =W_k+ S_{k, k-2}, \; 
X_k^{(k+1)} =W_k+ S_{k, k+2}.
\ee
The \achved rate is $\rx =2$.

\tbf{Recovery.} 
For \user{k}, the \rcvd \msgs  are 
$X_{k-1}=(X_{k-1}^{(k)}, X_{k-1}^{(k-2)}     )   $ and $X_{k+1}=(X_{k+1}^{(k)}, X_{k+1}^{(k+2)}     )   $ with \compos
$X_{k-1}^{(k)}=W_{k-1}+ S_{k-1,k+1}  $,
$X_{k-1}^{(k-2)}=W_{k-1}+ S_{k-1,k-3}$, 
$X_{k+1}^{(k)}=W_{k+1}+ S_{k+1,k-1}$,  and
$X_{k+1}^{(k+2)}=W_{k+1}+ S_{k+1,k+3}  $. 
The desired input sum is recovered via
\begin{align}
\label{eq:recov at user k,gen schm}
X_{k-1}^{(k)}+X_{k+1}^{(k)} 
& \overset{\eqref{eq:msg Xk,gen schm},\eqref{eq:msg Xk compos,gen schm} }{=}
W_{k-1}+ S_{k-1,k+1}  + W_{k+1}+ S_{k+1,k-1} \notag\\
&  = W_{k-1}+W_{k+1}.
\end{align}

\tbf{Proof of \Secty.} Consider any \user{k}. It should not infer anything  about $(W_{k-1}, W_{k+1})$ beyond $W_{k-1}+ W_{k+1}$ from $X_{k-1}$ and $X_{k+1}$. 
Since the two  useful \msg \compos  
$X_{k-1}^{(k)}=W_{k-1}+ S_{k-1,k+1}  $ and 
$X_{k+1}^{(k)}=W_{k+1}+ S_{k+1,k-1}$ 
expose only the desired sum  $W_{k-1}+ W_{k+1}$,
we need to  ensure the two remaining \compos 
$X_{k-1}^{(k-2)}=W_{k-1}+ S_{k-1,k-3}$ and 
$X_{k+1}^{(k+2)}=W_{k+1}+ S_{k+1,k+3}$ leak no \info about $(W_{k-1}, W_{k+1})$.
This is  true \bcuz the keys $S_{k-1, k+1},  S_{k-1,k-3},S_{k+1,k+3}  \notin  \Zc_k       $ are mutually \indep and unknown to \user{k}.
More  rigorously, we have
\begin{subequations}
\begin{align}
& I\left(X_{k-1}, X_{k+1}; W_{k-1}, W_{k+1}| W_{k-1}+W_{k+1}, W_k,\Zc_k \right)  \notag \\
&  \overset{\eqref{eq:key-key,key-input indepce}}{=}  I\left(X_{k-1}, X_{k+1}; W_{k-1}, W_{k+1}| W_{k-1}+W_{k+1},\Zc_k \right)
\label{step0}
\\
& = H\left(X_{k-1}, X_{k+1}   | W_{k-1}+ W_{k+1} , \Zc_k  \right)  \notag\\
& \quad - H\left( X_{k-1}, X_{k+1} |W_{k-1}, W_{k+1} , \Zc_k  \right)
\label{step1}
\end{align}
\end{subequations}
The second term in \eqref{step1} is equal to 
\begin{subequations}
\begin{align}
& H\left( X_{k-1}, X_{k+1}  |W_{k-1}, W_{k+1}, \Zc_k  \right)  \notag\\
&  \overset{\eqref{eq:msg Xk,gen schm}, \eqref{eq:msg Xk compos,gen schm} }{=}
 H\left(S_{k-1, k+1}, S_{k-1, k-3},   S_{k+1, k+3} |W_{\{k-1, k+1\}}, \Zc_k  \right)  \notag
 \\
 & \overset{\eqref{eq:key-key,key-input indepce}}{=}
  H\left(S_{k-1, k+1}, S_{k-1, k-3},   S_{k+1, k+3} | \Zc_k \right)
\label{stept22}
\\
& \overset{\eqref{eq:key-key,key-input indepce}}{=}
  H(S_{k-1, k+1}) +  H(S_{k-1, k-3}  )  + H (S_{k+1, k+3}  )=3, \notag
\end{align}
\end{subequations}
where  \eqref{stept22} is due to the \indepce  among  the keys and inputs. 
The first term  can be upper bounded by
\begin{subequations}
\begin{align}
& H\left( X_{k-1},  X_{k+1}    | W_{k-1} + W_{k+1}, \Zc_k  \right) \notag \\
&  \overset{\eqref{eq:msg Xk,gen schm},\eqref{eq:msg Xk compos,gen schm}, \eqref{eq:key-key,key-input indepce}  } {=}   H\left( X_{k-1},  X_{k+1} | W_{k-1} + W_{k+1}  \right) \label{step10}
\\
&= H\left( X_{\{k-1,k+1\}}, W_{k-1} + W_{k+1}  \right) -
H\left( W_{k-1} + W_{k+1} \right)  \notag
\\
& \overset{\eqref{eq:msg Xk compos,gen schm}}{=}
 H\big(
 X_{k-1}^{(k)},      X_{k-1}^{(k-2)},
  X_{k+1}^{(k)},      X_{k+1}^{(k+2)}, W_{k-1} + W_{k+1} \big) - 1  \notag\\
 & \overset{\eqref{eq:recov at user k,gen schm}}{=}
 H\big(  X_{k-1}^{(k)},      X_{k-1}^{(k-2)},
  X_{k+1}^{(k)},      X_{k+1}^{(k+2)} \big) - 1
\label{step11} \\
&  \le 3, \label{step12}
\end{align}
\end{subequations}
where \eqref{step10}  is due to the \indepce of the keys and inputs (see \eqref{eq:key-key,key-input indepce})  and  also  the fact that $S_{k-1, k+1}, S_{k-1, k-3 }, S_{k+1, k+3}$, on which $X_{k-1}  $ and $X_{k+1 } $ depend are not contained in $\Zc_k$; \eqref{step11} is due to the \recovy condition \eqref{eq:recov at user k,gen schm} at \user{k}; \eqref{step12} is \bcuz  each of the four \msg \compos contains one  bit and uniform  \distn maximizes the entropy.

\Aar,  $\eqref{step1}\le0$. Since \muinfo cannot be negative, we have  $\eqref{step1}=0$, proving \secty regarding \user{k}.

\section{Converse}
\label{sec:converse}

For \secagg over rings, a straightforward cut-set bound gives $\rx \ge 1$. This is \bcuz  $W_k$ appears  only at \user{k} and has to be recovered (in the form of $W_k +W_{k+2}$) at  \user{k+1},   a cut between users $k$  and $k+1$ implies $ H(X_k) \ge H(W_k) = L $, yielding $\rx\ge 1$ (see formal proof in \cite[Lemma 1]{zhang2026information}).
This cut-set bound is tight for $K=3,4$ but 
turns out  to be  loose when $K\ge 5$. 
In this  section, we derive a novel converse showing  $\rxstar \ge2 , \forall K\ge 5$.
Together with the  \achvb \schms, we fully \charz  $\rxstar$  for all $K$.
\begin{figure}[t]
    \centering
    \includegraphics[width=0.3\linewidth, height=0.27\linewidth]{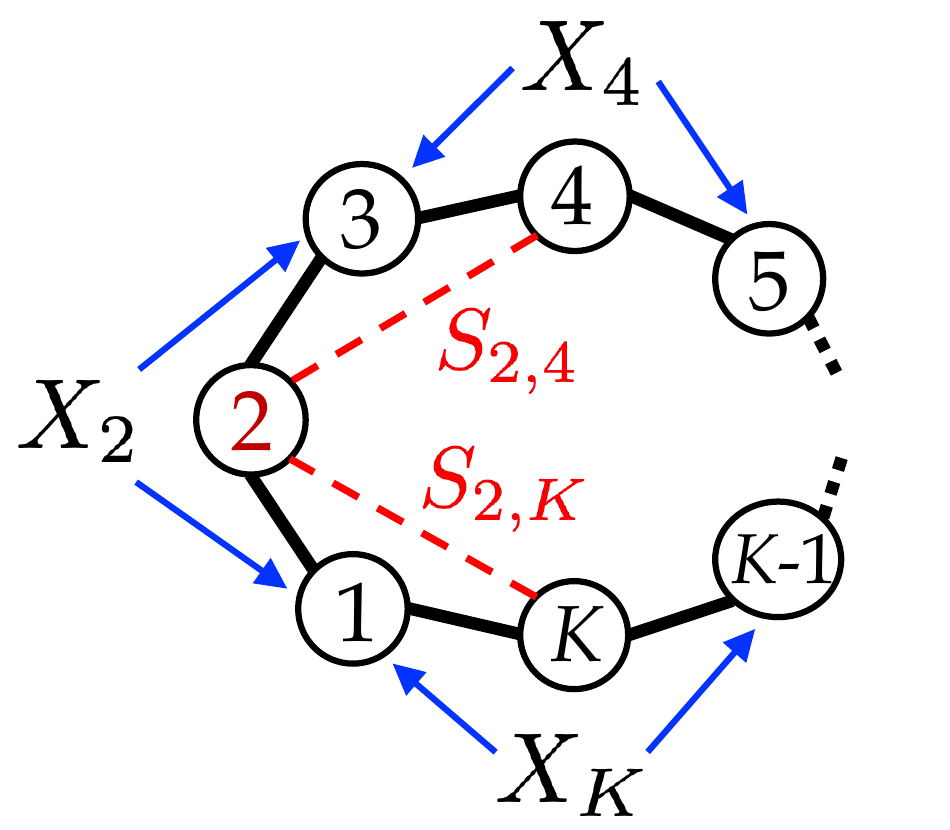}
    \caption{\small  $X_2$  supports  the \recovy of $W_2+W_K$ at \user{1},  and $W_2+W_4$ at \user{3}. The \pks  $S_{2,K}$ and $S_{1,2}$ are the shared randomness between $(X_2,X_K)$ and $(X_2,X_4)$, \resp. }
    \label{fig:ring segment,cvs}
\end{figure}

\tit{Proof outline:}
For $K\geq 5$, we prove that user $2$'s message must satisfy
$H(X_2)\geq 2L$ to support the recovery constraints of users 1 and 3 (see Fig. \ref{fig:ring segment,cvs}).
The key step is to identify from $X_2$ two components,
$U_2^{(1)}$ and $U_2^{(3)}$, which are respectively necessary for the
recoveries at users $1$ and $3$ and satisfy
$H(U_2^{(1)})\geq L$ and $H(U_2^{(3)})\geq L$.
We further show that
under the appropriate conditioning these two components are independent,
which implies $H(X_2)\geq H(U_2^{(1)})+H(U_2^{(3)})\geq 2L
$ and thus $\rx\ge 2$.

\subsubsection{\Recovy at \User{1}} 
Let  $\mrm{Dec}_k$ denote \user{k}'s decoder which takes in $X_{k-1},X_{k+1},W_k,\Zc_k$ as input and outputs $W_{k-1} + W_{k+1},k\in [K]$. 
For \user{1}, we have 
\begin{align}
    \mathrm{Dec}_1(X_2,X_K,W_1,\Zc_1)=W_2+W_K. 
\end{align}
Let $\Xc_2$ denote the alphabet of $X_2$. 
For each realization $(w_2, s_{1,2})$ of $(W_2,S_{1,2})$,  we define an equivalence relation 
$\sim_{1}^{(w_2,s_{1,2})}$ on $\Xc_2$ \af:
For any  $x_2,x_2' \in \Xc_2$, we say they are \eqvlt, denoted by
$x_2\sim_{1}^{(w_2,s_{1,2})} x_2'$,  if they produce the same decoder output, \ie, 
\begin{align}
    \mathrm{Dec}_1(x_2,x_K,w_1,\bm{z}_1)-w_2=\mathrm{Dec}_1(x_2',x_K,w_1,\bm{z}_1)-w_2, \label{eq:C-equi}
\end{align}
for all admissible realization  $(x_K,w_1,\bm{z}_1)$ of $(X_K,W_1,\Zc_1)$, where $S_{1,2}$ is fixed to $s_{1,2}$ in $\bm{z}_1$. 
Define  $U_2^{(1)}$ 
as the equivalence class of $X_2$,
conditioned on $(W_2,S_{1,2})$, under the equivalence relation induced by user 1's decoder:
\begin{align}
\label{eq:U21 def}
U_2^{(1)}\eqdef[X_2]_{\sim_{1}^{(W_2,S_{1,2})}},
\end{align}
Intuitively, $ U_2^{(1)}$ removes the decoder-irrelevant randomness caused by $\Zc_2\bksl \{S_{1,2}, S_{2,K}\}$
in $X_2$ and retains only the part of $X_2$ that affects user 1's reconstruction of the desired sum $W_2+W_K$.
$ U_2^{(1)}$ thus satisfies  
\begin{align} 
& H\big(W_K|U_2^{(1)},X_K,W_1,\Zc_1,W_2\big)=0,\label{eq:C-Rec} \\
& H \big(U_2^{(1)}|W_2,S_{1,2},S_{2,K}\big)=0,\label{eq:C-func}
\end{align}
where \eqref{eq:C-Rec} is due to  the recovery constraint at \user{1}, and
\eqref{eq:C-func} shows $U_2^{(1)}$ depends on $\Zc_2$ \tit{only} through $(S_{1,2},S_{2,K})$ but not  the remaining keys $\Zc_2\bksl \{S_{1,2}, S_{2,K}\}$.

\tbf{Proof of \eqref{eq:C-Rec}.}
By definition of the equivalence relation \eqref{eq:C-equi}, all realizations of $X_2$ in each  equivalence class (\ie, a realization of $\Utwoone$) produce the same  $W_K$ given $W_2$. Therefore, $\Utwoone$ preserves all the \nece  \info for decoding at \user{1} and  \eqref{eq:C-Rec} follows naturally.

\tbf{Proof of \eqref{eq:C-func}.}
\eqref{eq:C-func} is a key step in the converse proof: it removes the
irrelevant effect of the keys in
$\Zc_2\bksl \{S_{1,2}, S_{2,K}\}$ on $X_2$ through the quotient variable $\Utwoone$.
In particular, we show that  $\Utwoone$ depends only on two keys  $(S_{1,2}, S_{2,K})$.
Fix  $(W_2, S_{1,2}, S_{2, K})$ as $(w_2, s_{1,2}, s_{2, K})$. 
The side
information available to decoder $\mathrm{Dec}_1$, namely
$(X_K,W_1,\Zc_1)$, is independent of
$\mathcal Z_2\setminus\{S_{1,2},S_{2,K}\}=\{S_{2,k}\}_{k\in[3:K-1]}$.
Moreover, 
the decoded value  $W_K$ in \eqref{eq:C-equi} is also \indep  of $\{S_{2,k}\}_{k\in  [3:K-1]} $.
Therefore, any two realizations of $X_2$ induced solely by different
values of $\{S_{2,k}\}_{k\in[3:K-1]}$, while keeping
$(w_2,s_{1,2},s_{2,K})$ fixed, are indistinguishable from
$\mathrm{Dec}_1$'s viewpoint, \ie,  they belong to the same
equivalence class. Hence the equivalence class representative
$U_2^{(1)}$ is fully determined by $(W_2,S_{1,2},S_{2,K})$, which gives $H(U_2^{(1)}|W_2,S_{1,2},S_{2,K})=0$.
\hfill $\blacksquare$

\Inadd, the \secty  constraint  at \user{1} requires
\be 
\label{eq:C-Sec} 
H(W_K|X_K,W_1,\Zc_1,W_2) =  L. 
\ee 

\tbf{Proof of \eqref{eq:C-Sec}.}
Denote $C\eqdef (W_2,W_1, \Zc_1)$.
\Secty at \user{1} requires 
$I(X_2,X_K ;W_2 ,W_K|W_2+W_K, W_1, \Zc_1)=0$ (see \eqref{eq:security}), which implies $0=I(X_K ;W_2 ,W_K|W_2+W_K, W_1, \Zc_1)= I(X_K ;W_2 |W_2+W_K, W_1, \Zc_1) + I(X_K;W_K|W_K, C)$, yielding
\be
\label{eq:I(XK;WK)=0}
I(X_K;W_K|W_K, C)=0.
\ee 
Thus, 
\begin{subequations}
\begin{align}
 H(W_K|X_K,C) & \overset{\eqref{eq:I(XK;WK)=0}
}{=} H(W_K|X_K,C) +I(X_K;W_K|W_K, C) \notag \\
&= H(W_K|W_1, \Zc_1,W_2)\\
& \overset{\eqref{eq:input indep},\eqref{eq:key-key,key-input indepce}}{=}  H(W_K)=L, \label{eq:step00}     
\end{align} 
\end{subequations}
where \eqref{eq:step00} follows from the independence among the inputs and the keys.
\hfill $\blacksquare$

Hence, we have
\begin{subequations}
\begin{align}
L & \overset{\eqref{eq:C-Sec}}{=} H(W_K|X_K,W_1,\Zc_1,W_2)  \\
& =I\big(U_2^{(1)};W_K|X_K,W_1,\Zc_1,W_2\big) \notag\\ & \quad\;+ H(W_K|U_2^{(1)},X_K,W_1,\Zc_1,W_2)\\
&  \overset{\eqref{eq:C-Rec}}{=} 
I\big(U_2^{(1)};W_K|X_K,W_1,\Zc_1,W_2\big)\\
& \le H\big(U_2^{(1)}|X_K,W_1,\Zc_1,W_2\big)\\
& \le H\big(U_2^{(1)}|W_1,\Zc_1,W_2\big) \label{step0u21}\\
& \overset{\eqref{eq:C-func}, \eqref{eq:key-key,key-input indepce}  }{=}
H\big(U_2^{(1)}|W_2,S_{1,2}\big), \label{step0u22} \\
\Rightarrow & \;   H\big(U_2^{(1)}|W_2,S_{1,2}\big) \ge  L, \label{eq:u21lb}
\end{align}
\end{subequations}
where \eqref{step0u22} is \bcuz  $\Utwoone$ is \indep of $W_1$, and is completely determined by $(W_2, S_{1,2},S_{2,K})$ (see \eqref{eq:C-func}) where $\{W_2, \Zc_1 \}\cap \{W_2, S_{1,2},S_{2,K}\} =\{W_2, S_{1,2}\}$.

\subsubsection{\Recovy at \User{3}}

Similarly, regarding the  \recovy of $W_2 +W_4 $ at \user{3}, 
we can define another \eqvlce relation 
$\sim_{3}^{(W_2,S_{2,3})}$ on \user{3}'s decoder, and define 
\be 
\label{eq:U23 def}
U_2^{(3)}\eqdef[X_2]_{\sim_{3}^{(W_2,S_{2,3})}}
\ee 
as the equivalence class of $X_2$,
conditioned on $(W_2,S_{2,3})$, under the equivalence relation $\sim_{3}^{(W_2,S_{2,3})}$. 
We have
\begin{align}
&H\big(W_4|U_2^{(3)},X_4,W_3,\Zc_3,W_2\big)=0,\\
& H\big(U_2^{(3)}|W_2,S_{2,3},S_{2,4}\big)=0,\label{eq:U23 Func}\\
&H(W_4|X_4,W_3,\Zc_3,W_2)=L,
\end{align}
which together yields 
\be
H\big(U_2^{(3)}|W_2, S_{2,3}\big)\ge L. \label{eq:u23lb}
\ee

\subsubsection{Merging  $U_2^{(1)}$ and $U_2^{(3)}$ } 

By \eqref{eq:C-func} and \eqref{eq:U23 Func}, 
 $U_2^{(1)}$  is determined by
$(W_2, S_{1,2}, S_{2,K})$, and $U_2^{(3)}$  is determined by
$(W_2, S_{2,3}, S_{2,4})$. \Thf, given $(W_2, S_{1,2}, S_{2,3})$, $U_2^{(1)}$ depends only on $S_{2,K}$ and $U_2^{(3)}$ depends only on $S_{2,4}$. Since $S_{2,K}$ is \indep of $S_{2,4}$, we have the conditional \indepce
\be 
\label{eq:cond. indepce of U21,U23}
I\big(U_2^{(1)}; U_2^{(3)}| W_2, S_{1,2}, S_{2,3}    \big)=0.
\ee 

Now, we derive a lower bound on $H(X_2)$ \af:
\begin{subequations}
\begin{align}
& H(X_2 )   \ge  H(X_2|W_2 )\\
& \; \ge H(X_2|W_2, S_{1,2}, S_{2,3} )\\
& \; \overset{\eqref{eq:U21 def}, \eqref{eq:U23 def} }{=}   H\big(X_2, U_2^{(1)},U_2^{(3)}  |W_2, S_{1,2}, S_{2,3} \big) \label{eq:step0,H(X2) proof}\\
& \; \ge H\big(U_2^{(1)},U_2^{(3)}  |W_2, S_{1,2}, S_{2,3} \big)\\
& \overset{\eqref{eq:cond. indepce of U21,U23}}{=}
H\big(U_2^{(1)} |W_2, S_{1,2}, S_{2,3} \big)
+ H\big(U_2^{(3)} |W_2, S_{1,2}, S_{2,3} \big)\label{eq:step1,H(X2) proof}
\\
&  \overset{\eqref{eq:C-func}, \eqref{eq:U23 Func}}{=}
H\big(U_2^{(1)} |W_2, S_{1,2} \big)
+ H\big(U_2^{(3)} |W_2, S_{2,3} \big)\label{eq:step15,H(X2) proof}
\\
 & \; \overset{\eqref{eq:u21lb},\eqref{eq:u23lb}}{\ge  } 2L, \label{eq:step2,H(X2) proof}
 \\
 & \Rightarrow  \rx \eqdef  H(X_2)/L \ge  2. 
\end{align}
\end{subequations}
\eqref{eq:step0,H(X2) proof} is \bcuz $U_2^{(1)}$  and $U_2^{(3)}$ are by definition  functions of $(X_2,W_2, S_{1,2})$ and $(X_2,W_2, S_{2,3})$, \resp; \eqref{eq:step1,H(X2) proof} is due to the conditional  \indepce  between  $U_2^{(1)}$  and $U_2^{(3)}$ (see  \eqref{eq:cond. indepce of U21,U23}); 
\eqref{eq:step15,H(X2) proof} follows from
\eqref{eq:C-func},\eqref{eq:U23 Func}),
and pairwise key independence:
given \((W_2,S_{1,2})\), \(U_2^{(1)}\) depends only on \(S_{2,K}\),
which is independent of \(S_{2,3}\), and given
\((W_2,S_{2,3})\), \(U_2^{(3)}\) depends only on \(S_{2,4}\),
which is independent of \(S_{1,2}\).
In \eqref{eq:step2,H(X2) proof},  the lower bounds on $U_2^{(1)}$  and $U_2^{(3)}$ (see \eqref{eq:u21lb}, \eqref{eq:u23lb}) are invoked. 
\Aar, we have proved  $\rx \ge 2,\forall K \ge 5$.

\section{Conclusion}
\label{sec:conclusion}

This paper characterized the optimal per-user communication rate for topological secure aggregation over ring networks with pairwise keys. We showed a sharp transition: 
$\rxstar=1$ for $K \le 4$ and $\rxstar=2$ for all $K \ge 5$. A linear masking scheme using only distance-$2$ pairwise keys achieves these rates, while a matching converse proves optimality by exploiting recovery, security, and pairwise-key independence. The result shows that sparse topology and restricted key structure can jointly increase communication rate. Extending the characterization to general sparse graphs and user dropouts remains open.


\bibliographystyle{IEEEtran}
\bibliography{references_secagg.bib}

\end{document}